\documentclass[preprint,showpacs,preprintnumbers]{revtex4}
\usepackage{amsfonts}
\usepackage{amsmath}
\usepackage{amssymb}
\usepackage{graphicx}
\usepackage{fancyhdr}
\makeatletter
\newcommand{\bm}[1]{\mbox{\boldmath{$#1$}}}

\newcommand{\Rmnum}[1]{\expandafter\@slowromancap\romannumeral #1@}
\makeatother 

\begin{document}

\title{Derivation of exact master equation with stochastic description: Dissipative harmonic oscillator}
\author{Haifeng Li}
\affiliation{College of Chemistry, Beijing Normal University,
Beijing 100875, China}
\author{Jiushu Shao}\email{jiushu@bnu.edu.cn}
\affiliation{College of Chemistry, Beijing Normal University,
Beijing 100875, China}
\author{Shikuan Wang}
\affiliation{College of Chemistry, Beijing Normal University,
Beijing 100875, China}

\begin{abstract}
A systematic procedure for deriving the master equation of a
dissipative system is reported in the framework of stochastic
description. For the Caldeira-Leggett model of the
harmonic-oscillator bath, a detailed and elementary derivation of
the bath-induced stochastic field is presented.~The dynamics of the
system is thereby fully described by a stochastic differential
equation and the desired master equation would be acquired with
statistical averaging. It is shown that the existence of a
closed-form master equation depends on the specificity of the system
as well as the feature of the dissipation characterized by the
spectral density function. For a dissipative harmonic oscillator it
is observed that the correlation between the stochastic field due to
the bath and the system can be decoupled and the master equation
naturally comes out. Such an equation possesses the Lindblad form in
which time dependent coefficients are determined by a set of
integral equations. It is proved that the obtained master equation
is equivalent to the well-known Hu-Paz-Zhang equation based on the
path integral technique. The procedure is also used to obtain the
master equation of a dissipative harmonic oscillator in
time-dependent fields.
\end{abstract}
\date{\today}
\maketitle
\section{introduction}
In the real world physical systems are not exempt from the
disturbance of their surrounding environment. Because of dissipation
caused by system-environment
couplings~\cite{Leggett,Hanggi,Weiss,Breure,Ingold}, the dynamics of
open systems will change dramatically from reversible to
irreversible behavior. The most representative model of the
dissipative dynamics is the Brownian motion, a subject extensively
investigated in many academic areas ranging from natural sciences,
engineering to social sciences~\cite{C.W. Gardiner,M.F.M.
Osborne,J.A. Freund,M. Kijima,F. Beichelt}. As one observes that the
environment makes the dynamics of the system a stochastic process,
it is naturally to use random noises to imitate the influence of the
environment, which results in the phenomenological Langevin equation
and Fokker-Planck equation methods~\cite{H. Risken}. The Langevin
equation describes the motion of a classical Hamiltonian system
subjected additionally to a noise due to the environment, while the
Fokker-Planck equation depicts the evolution of the phase-space
probability for an ensemble of the system. The most successful
microscopic model for the environment consists of infinite number of
effective harmonic oscillators~\cite{V.B. Magalinskij}, which has
been shown to be a generic heat bath~\cite{Caldeira, Hanggi1}. The
harmonic-oscillator bath with linear couplings to the system under
study is called the Caldeira-Leggett model in the literature.
\par Traditionally dissipative dynamics is an essential subject of nonequilibrium processes and plays
an indispensable role in condensed-phase dynamics, transport as well
as spectroscopy~\cite{Nitzan,Grabert, C.H. Wang}. Nowadays the rapid
development of quantum information and quantum measurement provides
extraordinary paradigms for exploring dissipative effect, especially
the interplay of quantum coherence, dephasing and
relaxation~\cite{Goan, Blencowe, M.D. Ventra, H.G. Craighead}.
Because it is an inherent many-body problem, dissipative dynamics is
usually very difficult to solve exactly. There are, however, still
several general frameworks available for investigating the
diversified features of open systems~\cite{Weiss, Breure}. For
instance, the projection operator technique developed by
Nakajima~\cite{Nakajima} and Zwanzig~\cite{R.Zwanzig} has been
widely used in fields of spectroscopy and quantum optics where
dissipation in general is weak and thereby perturbation treatment
can be employed. The influence functional method based on the path
integral, on the other hand, which was first proposed by Feynman and
Vernon~\cite{Feynman} and popularized by Caldeira and
Leggett~\cite{Caldeira} has been shown a powerful tool for
theoretical analysis~\cite{Leggett}. This method has also been
implemented as a numerical technique to simulate dynamics of
dissipative two-state and three-state systems~\cite{C.H. Mark,N.
Makri}. Following Feynman, Stockburger {\em et al.} interpreted the
influence functional in the Caldeira and Leggett model as a random
field and put forward a stochastic formulation for the dissipative
dynamics~\cite{Stockburger}. Along this line one of the authors
extended the stochastic idea to general heat baths and was able to
establish that the influence of the heat bath on the system can be
fully characterized by its induced stochastic fields~\cite{Shao}.
Based on the stochastic description, a deterministic approach
comprising hierarchical equations of motion was developed~\cite{Shao
Hierarchical,Denis, Tanimura, Yan}. This scheme is only appliable to
specific dissipation where the bath-induced stochastic field is the
Ornstein-Ulenbeck type or the like. Fortunately it is shown, that
the hierarchy method combined with {\em stochastic realization} can
be used to accurately simulate zero-temperature dynamics of
dissipative two-state systems at strong dissipation~\cite{Shao
Hierarchical}.
\par Analytically solvable models are always desired
because they provide deep insights into the understanding of
dissipation and benchmark results for comparison when developing
approximations. It is unfortunate that in dissipative dynamics only
few systems are analytically solvable. Good examples include models
of the harmonic oscillator~\cite{Fleming,Dekker,F.Haake,P.Schramm,
Unruh,  V. Ambegaokar, Hu,Yu,H.Grabert,Connell,Calzetta,Strunz,
Chou,Chou1,Xu,peter,peter1,peter2}, the free
particle~\cite{Ambegaokar}, and the parabolic
barrier~\cite{Wolynes,H.Dekker,Ankerhold,Eli}. In this paper we will
show that the stochastic decoupling approach can be invoked to
obtain the master equation of the dissipative harmonic oscillator
with or without driving fields in a straightforward and elementary
way. Although some results were already reported and the method was
briefly outlined in previous work~\cite{Hu,Yu}, a full procedure by
which one can follow step-by-step is yet to supply. This paper will
fill this gap. We will give a detailed report on the derivation of
the master equation for dissipative linear systems from the
establishment of the stochastic differential equation of the
dissipative system to the statistical averaging that leads to the
desired master equation.
\par The paper is organized as follows.
In Sec.~\Rmnum{2} we briefly review the stochastic description of
quantum dissipation. In Sec.~\Rmnum{3} we apply the stochastic
formulation to a dissipative harmonic oscillator and derive its
exact master equation. In Sec.~\Rmnum{4} we prove the equivalence
between our results and the known results, Hu-Paz-Zhang master
equation. In Sec.~\Rmnum{5} we extend the scheme to the case of
the harmonic oscillator in time-dependent external fields and work
out the master equation. We conclude the paper in Sec.~\Rmnum{6}.
\section{Theory}
\subsection{Stochastic description: Primer}
To study the dissipative dynamics, we start with a system-plus-bath model defined by the Hamiltonian,
\begin{equation}
\hat{H} = \hat{H}_{s}+\hat{H}_{b}+\hat{H}_{int},
\end{equation}
where~$\hat{H}_{s}$~is the Hamiltonian of the renormalized system of
interest, $\hat{H}_{b}$~the Hamiltonian of the bath
and~$\hat{H}_{int}$~the interaction energy between the system and
the bath. Without loss of generality the couplings can be written
as~$\hat{H}_{int}=\sum_{\alpha}\hat{f}_{\alpha}\hat{g}_{\alpha}$,
where~$\hat{f}_{\alpha}$~are the operators of the system
and~$\hat{g}_{\alpha}$~the operators of the bath. According to
quantum mechanics the evolution of the total system obeys the
Liouville equation, namely,
\begin{equation}
i\hbar\frac{\partial \rho(t)}{\partial t}=\left[\hat{H},\rho(t)\right],
\end{equation}
where~$\rho(t)$~is the density matrix. Keep in mind that we are only
interested in the dynamics of the system. The reduced density
matrix~$\tilde{\rho}_{s}(t)=\textup{Tr}_{b}\rho(t)$~provides
sufficient information we require. Because the complexity of
dissipative dynamics lies in the coupling between the system and the
bath, it would be desired to decouple the interaction in such a way
that the evolution of the bath will no longer be explicitly involved in the evolution of the system.~Actually, it was shown by
one of the authors that the Hubbard-Stratonovich transformation can
be used to convert the system-bath coupling during the evolution to
the correlation of stochastic processes for the separated but random
system and bath~\cite{Shao}.~Doing this, the price one has to pay is
to introduce auxiliary stochastic fields in subsystems.~Later, it
was recognized that a simpler and natural language way to formulate
the dissipative dynamics as a stochastic one is the It\^o
calculus~\cite{Shao}.
\par The It\^o calculus is concerned with a Wiener
process~$W(t)=\int_{0}^{t}dt^{\prime}\mu(t^{\prime})$,
where~$\mu(t)$~is a Gaussian white noise with zero mean and delta
function correlation, i.e.,~$M\{\mu(t)\}=0$~and
~$M\left\{\mu(t)\mu(t^{\prime})\right\}=\delta(t-t^{\prime})$. Roughly speaking, the white noise~$\mu(t)$~can be regraded as a series of independent random numbers at time slices~$t_{0}=0, t_{1},..., t_{N}=t$, where one can simply assume the time steps~$\Delta t_{j}=t_{j}-t_{j-1}=\Delta t$~are uniform and~$\Delta t\rightarrow 0$. Thus the distribution function for any~$\mu_{j}=\mu(t_{j})$~is
\begin{equation*}
P\left(\mu_{j}\right)=\lim_{\Delta t\rightarrow 0}\sqrt{\frac{\Delta t}{2\pi}}e^{-\frac{\Delta t}{2}\mu_{j}^2}.
\end{equation*}
For a random process~$F\left[\mu\right]$~or~$F\left(\mu_{1},...,\mu_{N}\right)$~in discrete-time representation, the stochastic averaging~$M$~means the expectation value of~$F\left(\mu_{1},...,\mu_{N}\right)$, i.e.,
\begin{align*}
M\left\{F[\mu]\right\}=\int_{-\infty}^{\infty}\prod_{j=1}^{N}\left[d\mu_{j}P\left(\mu_{j}\right)\right]F\left(\mu_{1},...,\mu_{N}\right).
\end{align*}
Note that~$dW$~is of the order~$\sqrt{dt}$~and~$\left(dW\right)^{2}=dt$.
To apply the It\^o calculus we now consider the Liouville dynamics
Eq.~(2) with a disentangled initial
state~$\rho(0)=\rho_{s}(0)\rho_{b}(0)$. When calculating the time
derivative of a composite stochastic process constructed from the
Wiener process, one should take into account the second order
contribution of~$dW$. Let us analyze the following stochastic
differential equations,
\begin{align}
i\hbar
d\rho_{s}&=\left[\hat{H}_{s},\rho_{s}\right]dt+\frac{\sqrt{\hbar}}{2}\sum_{\alpha}\left[\hat{f}_{\alpha},\rho_{s}\right]dW_{1\alpha}
+i\frac{\sqrt{\hbar}}{2}\sum_{\alpha}\left\{\hat{f}_{\alpha},\rho_{s}\right\}dW^{*}_{2\alpha}\\
\intertext{and} i\hbar
d\rho_{b}&=\left[\hat{H}_{b},\rho_{b}\right]dt+\frac{\sqrt{\hbar}}{2}\sum_{\alpha}\left[\hat{g}_{\alpha},\rho_{b}\right]dW_{2\alpha}
+i\frac{\sqrt{\hbar}}{2}\sum_{\alpha}\left\{\hat{g}_{\alpha},\rho_{b}\right\}dW^{*}_{1\alpha},
\end{align}
where~$W_{1\alpha}(t)=\int_{0}^{t}dt^{\prime}\left[\mu_{1\alpha}(t^{\prime})+i\mu_{4\alpha}(t^{\prime})\right]$
and~$W_{2\alpha}(t)=\int_{0}^{t}dt^{\prime}\left[\mu_{2\alpha}(t^{\prime})+i\mu_{3\alpha}(t^{\prime})\right]$~are
complex Wiener processes with independent Gaussian white
noises~$\mu_{k\alpha}(t)~(k=1-4)$. Apparently there is no direct
interaction between the two subsystems and the equations of motion for
the system and the bath Eqs.~(3) and (4) are independent. Using
the It\^o calculus of complex Wiener
process,~$dW_{\alpha}dW_{k}=dW_{\alpha}^{*}dW_{k}^{*}=0$~and
$dW_{\alpha}dW_{k}=2\delta_{\alpha
k}dt$, and nonanticipating
property, that is, the combined stochastic process~$\rho_{s}(t)\rho_{b}(t)$~is statistically independent of~$dW_{1(2),t}$, $M\{\rho_{s}(t)\rho_{b}(t)dW_{1(2),t}\}=0$,
we can readily prove
\begin{align}
i\hbar
d\left(M\left\{\rho_{s}(t)\rho_{b}(t)\right\}\right)&=i\hbar M\left\{d\rho_{s}(t)\rho_{b}(t)+\rho_{s}(t)d\rho_{b}(t)+d\rho_{s}(t)d\rho_{b}(t)\right\}\nonumber\\
&=\left[\hat{H}_{s}+\hat{H}_{b}
+\sum_{\alpha}\hat{f}_{\alpha}\hat{g}_{\alpha},M\left\{\rho_{s}(t)\rho_{b}(t)\right\}\right]dt.
\end{align}
Therefore, $M\{\rho_{s}(t)\rho_{b}(t)\}$~satisfies the Liouville
equation Eq.~(2) and is of course identical with the density matrix
$\rho(t)$~of the total system.
\par With the help of the It\^{o}
calculus, therefore, we are able to illustrate how the interaction
between the system and bath is decoupled, as shown in Eqs.~(3) and
(4). As a consequence, the system as well as the bath is subjected
to random fields. Again, it should be stressed that we want to
calculate the reduced density matrix,
$\tilde{\rho}_{s}(t)=\textup{Tr}_{b}M\{\rho_{s}(t)\rho_{b}(t)\}$. To
this end we change the operation order of the tracing over the
degrees of freedom of the bath and the stochastic averaging to
obtain
\begin{equation}
\tilde{\rho}_{s}(t)=M\{\rho_{s}(t)\textup{Tr}_{b}\rho_{b}(t)\}.
\end{equation}
The formal expression of~$\textup{Tr}_{b}\rho_{b}(t)$ can be
feasibly acquired by solving its equation of motion Eq.~(4),
\begin{equation}
\textup{Tr}_{b}\rho_{b}(t)=\exp\left\{\frac{1}{\sqrt{\hbar}}\sum_{\alpha}\int_{0}^{t}dt^{\prime}\left[\mu_{1\alpha}(t^{\prime})
-i\mu_{4\alpha}(t^{\prime})\right]\bar{g}_{\alpha}(t^{\prime})\right\},
\end{equation}
where introduced are the bath-induced stochastic fields
\begin{equation}
\bar{g}_{\alpha}(t)=\textup{Tr}_{b}\left\{\hat{g}_{\alpha}\bar{\rho}_{b}(t)\right\}
\end{equation}
with~$\bar{\rho}_{b}(t)=\rho_{b}(t)/\textup{Tr}_{b}\rho_{b}(t)$ the
normalized density matrix for the random bath. Now comes a crucial
step. We can change the measure of stochastic processes to absorb the trace of the bath, that is, we modify the Wiener process~$W_{1\alpha}(t)$
as~$W_{1\alpha}(t)+2\int_{0}^{t}dt^{\prime}\bar{g}_{\alpha}(t^{\prime})/\sqrt{\hbar}$.
Complying with this,~$\rho_{s}(t)$~will also change accordingly. The
mathematical manipulation is nothing but the Girsanov
transformation~\cite{ksendal,D. Gatarek}. To illustrate this transformation clearly, we again resort to the discrete-time representation.
Now inserting Eq.~(7) into Eq.~(6), one has
\begin{equation}
\tilde{\rho}_{s}(t)=\int_{-\infty}^{\infty}\prod_{\alpha}\prod_{k=1}^{4}\prod_{j}^{N}d\mu_{j,k\alpha}P\left(\mu_{j,k\alpha}\right)\exp\left\{
\frac{\Delta t}{\sqrt{\hbar}}\sum_{\alpha}\sum_{j=1}^{N}\left[\mu_{j,1\alpha}-i\mu_{j,4\alpha}\right]\bar{g}_{\alpha}(t_{j})\right\}
\rho_{s}\left(t,\left\{\mu_{j,k\alpha}\right\}\right).
\end{equation}
Note that in this expression one can put the contribution due to the bath-induced field into the distribution function~$P\left(\mu_{j,1(4)\alpha}\right)$, namely,
\begin{align*}
&\prod_{\alpha}P(\mu_{j,1\alpha})P(\mu_{j,4\alpha})\exp\left\{\frac{\Delta t}{\sqrt{\hbar}}\sum_{\alpha}\left[\mu_{j,1\alpha}-i\mu_{j,4\alpha}\right]\bar{g}_{\alpha}(t_{j})\right\}
\\&=\prod_{\alpha}P\left(\mu_{j,1\alpha}-\frac{1}
{\sqrt{\hbar}}\bar{g}_{\alpha}(t_{j})\right)P\left(\mu_{j,4\alpha}+\frac{i}{\sqrt{\hbar}}\bar{g}_{\alpha}(t_{j})\right).
\end{align*}
Now define new variables~$
\tilde{\mu}_{j,1\alpha}=\mu_{j,1\alpha}-\bar{g}_{\alpha}(t_{j})/\sqrt{\hbar}$,
$\tilde{\mu}_{j,2\alpha}=\mu_{j,2\alpha}$,
$\tilde{\mu}_{j,3\alpha}=\mu_{j,3\alpha}$, and~$\tilde{\mu}_{j,4\alpha}=\mu_{j,4\alpha}+i\bar{g}_{\alpha}(t_{j})/\sqrt{\hbar}$.
Because both~$\mu_{j,1\alpha}$~and~$\mu_{j,4\alpha}$~are independent of~$\bar{g}_{\alpha}(t_{j})$, the Jacobian of the variable change is equal to one. With these new variables, therefore, Eq.~(9) becomes
\begin{align*}
\tilde{\rho}_{s}(t)&=\int_{-\infty}^{\infty}\prod_{\alpha}\prod_{k=1}^{4}\prod_{j=1}^{N}d\tilde{\mu}_{j,k\alpha}P\left(\tilde{\mu}_{j,k\alpha}\right)
\rho_{s}\left(t,\left\{\tilde{\mu}_{j,1\alpha}+\frac{1}{\sqrt{\hbar}}\bar{g}_{\alpha}(t_{j}),\tilde{\mu}_{j,2\alpha},\tilde{\mu}_{j,3\alpha},
\tilde{\mu}_{j,4\alpha}-\frac{i}{\sqrt{\hbar}}\bar{g}_{\alpha}(t_{j})\right\}\right)
\\&=M\left\{\rho_{s}\left[t,\left\{\tilde{\mu}_{k\alpha}\right\}\right]\right\},
\end{align*}
where the second equality results from the continuous representation. Notice that~$\rho_{s}(t)\equiv\rho_{s}\left[t,\left\{\mu_{k\alpha}\right\}\right]$~satisfies Eq.~(3). Changing the underlying random processes in Eq.~(3) from~$\left\{\mu_{k\alpha}\right\}$~to~$\left\{\tilde{\mu}_{k\alpha}\right\}$, one obtains the equation for the {\em new} density matrix~$\rho_{s}(t)\equiv\rho_{s}\left[t,\left\{\tilde{\mu}_{k\alpha}\right\}\right]$,
\begin{equation}
i\hbar d\rho_{s} =
\left[\hat{H}_{s}+\sum_{\alpha}\overline{g}_{\alpha}(t)\hat{f}_{\alpha},\rho_{s}\right]dt
+\frac{\sqrt{\hbar}}{2}\sum_{\alpha}\left[\hat{f}_{\alpha},\rho_{s}\right]dW_{1\alpha}
+i\frac{\sqrt{\hbar}}{2}\sum_{\alpha}\left\{\hat{f}_{\alpha},\rho_{s}\right\}dW^{*}_{2\alpha}.
\end{equation}
For brevity, here and in the following, the functional dependence of the random density matrix~$\rho_{s}(t)$~on~$\left\{\tilde{\mu}_{k\alpha}\right\}$~will not indicated explicitly.\par
Given~$\bar{g}_{\alpha}(t)$, one only needs to solve stochastic
differential equation Eq.~(10) and calculate the random average to obtain the
exact reduced density matrix~$\tilde{\rho}_{s}(t)$.
The bath-induced field in general evoke fast motion and thus~$\rho_{s}(t)$~as well as the white noises to which~$\rho_{s}(t)$~is subjected displays a smaller time scale than its average~$\tilde{\rho}_{s}(t)$. In the following
section we will consider the Caldeira-Leggett (CL) model for which
the bath induced field can be worked out explicitly. We would like
to emphasize that in the CL model there is only one interaction
term, while within the stochastic description one can deal with many
interaction terms.
\subsection{Caldeira-Leggett Model}
The Hamiltonian of a CL model reads
\begin{equation}
\hat{H}=
\frac{\hat{p}^2}{2M}+V(\hat{x})+\frac{1}{2}\sum_{j=1}^{N}\left[\frac{\hat{p}_{j}^{2}}{m_{j}}+m_{j}\omega^2_{j}\left(\hat{x}_{j}
+\frac{c_{j}}{m_{j}\omega_{j}^{2}}\hat{x}\right)^{2}\right],
\end{equation}
where the first two terms define the Hamiltonian of the system and
the third one induces the Hamiltonian of the bath, the interaction,
and a counter-term. The latter can directly be read off
as~\begin{equation*}
V^{\prime}(\hat{x})\equiv\frac{1}{2}M\hat{x}^{2}\tilde{\omega}^{2},
\end{equation*} where
\begin{equation}
\tilde{\omega}^{2}=\sum_{j}\frac{c_{j}^{2}}{Mm_{j}\omega^{2}_{j}}=\frac{2}{M\pi}\int_{0}^{\infty}d\omega
\frac{J(\omega)}{\omega}.
\end{equation}
$J(\omega)$~is the spectral density function
\begin{equation}
J(\omega)=\frac{\pi}{2}\sum_{j}\left[\frac{c_{j}^2}{m_{j}\omega_{j}}\delta(\omega-\omega_{j})\right],
\end{equation}
which exactly characterizes the effect of the environment. In
general the system-bath linear coupling not only provide a
dissipation mechanism for the system but also tend to renormalize
the potential~$V(\hat{x})$. To compensate for this renormalization
effect the counter-term is introduced. It ensures that the system
cannot lower its potential energy below the original uncoupled
value~\cite{Caldeira}.
\par Assume that the initial state of the bath is in a thermal equilibrium state for the
non-interacting harmonic oscillators, that is,
$\rho_{b}(0)=e^{-\beta\hat{H}_{b}}/\textup{Tr}_{b}\{e^{-\beta\hat{H}_{b}}\}$.
Because of the linearity, the dynamics of the random bath described
by Eq.~(4) is analytically solvable. Moreover, since there are no
interactions among the bath modes, the evolution of the bath is
fully determined by that of the individual mode. In other words, the
solution of Eq.~(4) can be written as
\begin{equation}
\rho_{b}(t)=\prod_{j}u_{j,1}(t,0)\rho_{b}(0)\prod_{j}u_{j,2}(0,t),
\end{equation}
where~$u_{j,1}(t,0)$~is the forward propagator of the
mode~$j$~dictated by the Hamiltonian
\begin{equation*}
\hat{h}_{1,j}(t)=\left(\frac{\hat{p}_{j}^{2}}{2m_{j}}+\frac{1}{2}m_{j}\hat{x}_{j}^{2}\omega_{j}^{2}\right)
+\frac{\sqrt{\hbar}}{2}\nu_{+}(t)c_{j}\hat{x}_{j}
\end{equation*}
and~$u_{j,2}(0,t)$~is the backward propagator dictated by
\begin{equation*}
\hat{h}_{2,j}(t)=\left(\frac{\hat{p}_{j}^{2}}{2m_{j}}+\frac{1}{2}m_{j}\hat{x}_{j}^{2}\omega_{j}^{2}\right)
+\frac{\sqrt{\hbar}}{2}\nu_{-}(t)c_{j}\hat{x}_{j}
\end{equation*}
with the stochastic fields~$\nu_{\pm}(t)=\mu_{2}(t)+i\mu_{3}(t)\pm
i\mu_{1}(t)\pm\mu_{4}(t)$. For a driven harmonic oscillator
\begin{equation}
\hat{h}(t)=\frac{\hat{p}^{2}}{2m}+\frac{1}{2}m\omega_{0}^{2}\hat{x}^{2}+\nu(t)\hat{x},
\end{equation}
its corresponding propagator~$u(t,0)$~can be worked out in terms of
the operator algebra method~\cite{Louisell}. To do this, it would be
better to work in the second quantization formalism. Now introduce
the creation and annihilation operators,
\begin{align}
a^{\dagger}=&\sqrt{\frac{m\omega_{0}}{2\hbar}}\hat{x}-\frac{i}{\sqrt{2\hbar
m\omega_{0}}}\hat{p}\\
\intertext{and}
a=&\sqrt{\frac{m\omega_{0}}{2\hbar}}\hat{x}+\frac{i}{\sqrt{2\hbar
m\omega_{0}}}\hat{p}.
\end{align}
From the commutation relation~$\left[\hat{x},\hat{p}\right]=i\hbar$,
one readily finds~$\left[a,a^{\dagger}\right]=1$. The Hamiltonian
Eq.~(15) then becomes
\begin{equation*}
\hat{h}(t)=\hat{h}_{0}+\sqrt{\frac{\hbar}{2m\omega_{0}}}\nu(t)\left(a+a^{\dagger}\right),
\end{equation*}
where~$\hat{h}_{0}=\hbar\omega_{0}\left(a^{\dagger}a+1/2\right)$~is
the Hamiltonian of an isolated harmonic oscillator. The
propagator~$u_{0}(t,0)$~dictated by~$\hat{h}_{0}$~has been well
known and the result can be found in several
textbooks~\cite{Feynman, Dittrich, Kleinert, Feynman1}. For the
driven case one uses the interaction representation to have
\begin{align*}
i\hbar\frac{\partial u_{I}(t,0)}{\partial
t}=\sqrt{\frac{\hbar}{2m\omega_{0}}}\nu(t)\left(ae^{-i\omega_{0}t}+a^{\dagger}e^{i\omega_{0}t}\right)u_{I}(t,0)
\end{align*}
with the initial condition~$u_{I}(0,0)=1$. Resorting to the
Baker-Campbell-Hausdorff formula~\cite{Baker}, we readily obtain
\begin{equation}
u_{I}(t,0)=\exp\left[d(t)a\right]\exp\left[e(t)a^{\dagger}\right]\exp[j(t)],
\end{equation}
where
\begin{align*}
d(t)&=-\frac{i}{\sqrt{2m\hbar\omega_{0}}}\int_{0}^{t}dt_{1}\exp\left(-i\omega_{0} t_{1}\right)\nu(t_{1}),\\
e(t)&=-\frac{i}{\sqrt{2m\hbar\omega}_{0}}\int_{0}^{t}dt_{1}\exp\left(i\omega_{0}
t_{1}\right)\nu(t_{1}), \intertext{and}
j(t)&=\frac{1}{2m\hbar\omega_{0}}\int_{0}^{t}dt_{1}\int_{0}^{t_{1}}dt_{2}\nu(t_{1})\nu(t_{2})\exp[i\omega_{0}(t_{1}-t_{2})].
\end{align*}
Upon using Eqs.~(16) and (17) and Baker-Campbell-Hausdorff formula
again, the propagator~$u_{I}(t,0)$~becomes
\begin{align*}
u_{I}(t,0)=&\exp\left\{\sqrt{\frac{m\omega_{0}}{2\hbar}}\left[d(t)+e(t)\right]\hat{x}\right\}\exp\left\{\frac{i}{\sqrt{2
m\hbar\omega_{0}}}\left[d(t)-e(t)\right]\hat{p}\right\}\exp\left\{\frac{1}{4}\left[d^{2}(t)-e^{2}(t)\right]\right\}.
\end{align*}
Finally with the given expressions
of~$u_{0}(t,0)$~and~$u_{I}(t,0)$~the propagator
$u(t,0)=u_{0}(t,0)u_{I}(t,0)$~is also available. Therefore, we
propose a straightforward procedure to
calculate~$u_{j,1}(t,0)$~and~$u_{j,2}(t,0)$. Inserting these results
into Eq.~(8) and calculating the trace we finally find
\begin{align}
\bar{g}(t)&=\sqrt{\hbar}\int_{0}^{t}dt^{\prime}\left\{\alpha_{R}(t-t^{\prime})\left[\mu_{1}(t^{\prime})-i\mu_{4}(t^{\prime})\right]
+\alpha_{I}(t-t^{\prime})\left[\mu_{2}(t^{\prime})+i\mu_{3}(t^{\prime})\right]\right\},
\end{align}
where~$\alpha_{R,I}(t)$~turn out to be the real and imaginary parts
of the autocorrelation function of the
``force''~$\hat{g}=\sum_{j}c_{j}\hat{x}_{j}$, namely,
\begin{equation}
\alpha(t)=\textup{Tr}_{b}\left\{\rho_{b}(0)\hat{g}(t)\hat{g}(0)\right\}=\frac{1}{\pi}\int_{0}^{\infty}d\omega
J(\omega)\left[\coth\left(\frac{\hbar\beta\omega}{2}\right)\cos\omega
t-i\sin\omega t\right]
\end{equation}
with~$\hat{g}(t)=e^{i\hat{H}_{b}t/\hbar}\hat{g}e^{-i\hat{H}_{b}t/\hbar}$.
One can see that~$\alpha_{R}(t)$~is dependent on temperature,
reflecting dissipation effects contributed from both quantum and
``classical'' motion, while~$\alpha_{I}(t)$~is independent of
temperature, embodying a pure quantum effect.
\subsection{Formal Solution}
As one can understand, we start from Eq.~(10), the working equation
in the framework of stochastic description. The formal solution of
this linear equation is~\cite{Shao}
\begin{equation}
\rho_{s}(t)=U_{1}(t,0)\rho_{s}(0)U_{2}(0,t),
\end{equation}
where~$U_{1}(t,0)$~is the forward propagator of the stochastic
system associated with the Hamiltonian
\begin{equation}
\hat{H}_{1}(t)=\hat{H}_{s}+\sum_{\alpha}\left[\bar{g}_{\alpha}(t)+\frac{\sqrt{\hbar}}{2}\eta_{+\alpha}(t)\right]\hat{f}_{\alpha}
\end{equation}
while~$U_{2}(0,t)$~is the backward propagator associated with the
Hamiltonian
\begin{equation}
\hat{H}_{2}(t)=\hat{H}_{s}+\sum_{\alpha}\left[\bar{g}_{\alpha}(t)+\frac{\sqrt{\hbar}}{2}
\eta_{-\alpha}(t)\right]\hat{f}_{\alpha}
\end{equation}
with~$\eta_{\pm\alpha}(t)=\mu_{1\alpha}(t)+i\mu_{4\alpha}(t)\pm
i\mu_{2\alpha}(t)\pm\mu_{3\alpha}(t)$. When we can take stochastic
average of random density matrix~$\rho_{s}(t)$~in principle,~we can
obtain the exact reduced density matrix~$\tilde{\rho}_{s}(t)$. To
derive the equation of motion for~$\tilde{\rho}_{s}(t)$~or the
master equation, however, it would be better to start from Eq.~(10).
Whenever the statistical average of the right hand side of Eq.~(10)
can be expressed explicitly in terms of~$\tilde{\rho}_{s}(t)$~and
other known operators of the system, one obtains a master equation.
\par To acquire the deterministic equation from the corresponding
stochastic differential equation for arbitrary noises is a
challenging, if not impossible task~\cite{A. Brissand, Kampen, V. I.
Klyatskin, M. Ban}. In the literature, one may find incorrect
results about the derivation of the deterministic equation for
Gaussian noises and similar statements in developing
master equation for open systems~\cite{Lujian}.
\par
For the Caldeira-Leggett model, upon substituting the expression of
the bath induced fields Eq.~(19) and carrying out the stochastic
averaging, Eq.~(10) becomes
\begin{align}
i\hbar \frac{d\tilde{\rho}_{s}(t)}{dt}=&\left[\hat{H}_{s},
\tilde{\rho}_{s}(t)\right]+\left[\hat{x},\int_{0}^{t}dt^{\prime}\alpha_{R}(t-t^{\prime})M\left\{\rho_{s,1}(t,t^{\prime})\right\}\right]\nonumber\\
&+\left[\hat{x},\int_{0}^{t}dt^{\prime}\alpha_{I}(t-t^{\prime})M\left\{\rho_{s,2}(t,t^{\prime})\right\}\right],
\end{align}
where~$\rho_{s,1}(t,t^{\prime})=\sqrt{\hbar}
\left[\mu_{1}(t^{\prime})-i\mu_{4}(t^{\prime})\right]\rho_{s}(t)$~and~$\rho_{s,2}(t,t^{\prime})=\sqrt{\hbar}\left[\mu_{2}(t^{\prime})
+i\mu_{3}(t^{\prime})\right]\rho_{s}(t)$. In the above derivation
the nonanticipating property of $\rho_{s}(t)$,
namely,~$M\{\rho_{s}(t)dW_{1(2)}(t)\}=0$~is used. It is obvious that
when there is no dissipation, only the first term remains on the
right hand side of Eq.~(24). Of course, this is the Liouville
equation of the system as it should be. As the stochastic fields are
added, the other two terms naturally come out. These terms lead to
the irreversibility caused by the coupling to the bath {\em \`{a}
la} its induced random fields. To obtain the expressions
of~$M\{\rho_{s,1(2)}(t,t^{\prime})\}$~we resort to the
Furutsu-Novikov theorem~\cite{Novikov}, that is,
\begin{equation}
M\left\{\mu(t^{\prime})F[\mu]\right\}=M\left\{\frac{\delta
F[\mu]}{\delta \mu(t^{\prime})}\right\}\end{equation} for a white
noise~$\mu(t)$~and its arbitrary functional~$F[\mu]$. Using this
theorem allows us to write down
\begin{equation*}
M\left\{\rho_{s,1}(t,t^{\prime})\right\}=\sqrt{\hbar}M\left\{\frac{\delta\rho_{s}(t)}{\delta\mu_{1}(t^{\prime})}-i\frac{\delta\rho_{s}(t)}{\delta\mu_{4}(t^{\prime})}\right\}\equiv
\hat{O}_{s,1}(t,t^{\prime})
\end{equation*}
and
\begin{equation*}
M\left\{\rho_{s,2}(t,t^{\prime})\right\}=\sqrt{\hbar}M\left\{\frac{\delta\rho_{s}(t)}{\delta\mu_{2}(t^{\prime})}+i\frac{\delta\rho_{s}(t)}{\delta\mu_{3}(t^{\prime})}\right\}\equiv
\hat{O}_{s,2}(t,t^{\prime}).
\end{equation*}
Therefore,~$M\left\{\rho_{s,1(2)}(t,t^{\prime})\right\}$~in
Eq.~(24)~can be replaced respectively
by~$\hat{O}_{s,1(2)}(t,t^{\prime})$. For brevity, they are called
the dissipation operators. With the formal solution
of~$\rho_{s}(t)$~Eq.~(21), one can directly work out the expression
of the functional derivatives
\begin{align}
\hat{O}_{s,1}(t,t^{\prime})=&-iM\left\{\hat{x}_{1}(t,t^{\prime})\rho_{s}(t)
-\rho_{s}(t)\hat{x}_{2}(t,t^{'})\right\}\\
\intertext{and}
\hat{O}_{s,2}(t,t^{\prime})=&M\left\{\hat{x}_{1}(t,t^{\prime})\rho_{s}(t)
+\rho_{s}(t)\hat{x}_{2}(t,t^{'})\right\},
\end{align}
where~$\hat{x}_{1}(t,t^{\prime})=U_{1}(t,t^{\prime})\hat{x}U_{1}(t^{\prime},t)$~and
$\hat{x}_{2}(t,t^{\prime})=U_{2}(t,t^{\prime})\hat{x}U_{2}(t^{\prime},t)$~are
the Heisenberg operators. To build up a master equation, of course,
we still need to find the expressions
of~$\hat{O}_{s,1(2)}(t,t^{\prime})$~in terms
of~$\tilde{\rho}_{s}(t)$~and other known operators. \par Some
comments are in orders. For arbitrary response
functions~$\alpha_{R(I)}(t)$, deriving the master equation relies on
whether the explicit expressions of dissipative
operators~$\hat{O}_{s,1(2)}(t,t^{\prime})$~can be worked out.
However, when~$\alpha_{R(I)}(t)$~are localized distribution
functions, say~$\alpha_{R(I)}(t)=\tilde{\alpha}_{R(I)}\delta(t)$,
Eq.~(24) becomes
\begin{equation}
i\hbar\frac{d\tilde{\rho}_{s}(t)}{dt}=\left[\hat{H}_{s},\tilde{\rho}_{s}(t)\right]+\left[\hat{x},\tilde{\alpha}_{R}M\{\rho_{s,1}(t,t)\}\right]
+\left[\hat{x},\tilde{\alpha}_{I}M\{\rho_{s,2}(t,t)\}\right].
\end{equation}
From Eqs.~(26) and (27), we
know~$M\left\{\rho_{s,1}(t,t)\right\}=-i\left[\hat{x},\tilde{\rho}_{s}(t)\right]$~and~$M\left\{\rho_{s,2}(t,t)\right\}=\left\{\hat{x},\tilde{\rho}_{s}(t)\right\}$.
Inserting into Eq.~(28) yields the master equation
\begin{equation}
i\hbar\frac{d\tilde{\rho}_{s}(t)}{dt}=\left[\hat{H}_{s},\tilde{\rho}_{s}(t)\right]-i\tilde{\alpha}_{R}\left[\hat{x},\left[\hat{x},\tilde{\rho}_{s}(t)\right]\right]
+\tilde{\alpha}_{I}\left[\hat{x},\left\{\hat{x},\tilde{\rho}_{s}(t)\right\}\right].
\end{equation}
Therefore, both the specificities of the system and the bath-induced
field determine the ``existence'' of the master equation.
\section{Derivation of Master Equation}
Now we shall derive the master equation of a dissipative harmonic oscillator
described by the Caldeira-Leggett model with
$V(\hat{x})=M\omega_{0}^{2}\hat{x}^{2}/2$. As discussed above, the
system-bath coupling modifies the potential of the system by
addition of a counter-term. In the present case the frequency of the
harmonic oscillator becomes a renormalized one,
\begin{equation*}
\omega^{2}=\omega_{0}^{2}+\tilde{\omega}^{2}.
\end{equation*}
To acquire the master equation we use the equation of motion in the
Heisenberg representation to find the
operators~$\hat{x}_{1(2)}(t,t^{\prime})$~in Eqs.~(26) and (27).
After some algebra we obtain
\begin{align*}
\hat{x}_{1(2)}(t,t^{\prime})=&\cos\omega(t-t^{\prime})\hat{x}
-\frac{\sin\omega(t-t^{\prime})}{M\omega}\hat{p}-\frac{1}{M\omega}\int_{t^{\prime}}^{t}dt_{1}\sin\omega(t_{1}-t^{\prime})
\left[\bar{g}(t_{1})+\frac{\sqrt{\hbar}}{2}\eta_{\pm}(t_{1})\right].\end{align*}
Inserting into Eqs.~(26)~and~(27) and carrying out the involved
stochastic averaging, we find
\begin{align}
 \hat{O}_{s,1}(t,t^{\prime})=&-i\cos\omega(t-t^{\prime})\left[\hat{x},\tilde{\rho}_{s}(t)\right]+\frac{i}{M\omega}
 \sin\omega(t-t^{\prime})\left[\hat{p},\tilde{\rho}_{s}(t)\right]\nonumber\\&
 +\frac{2}{M\omega}\int_{t^{\prime}}^{t}dt_{1}\int_{t_{1}}^{t}dt_{2}\sin\omega(t_{1}-t^{\prime})\alpha_{I}
 (t_{1}-t_{2})\hat{O}_{s,1}(t,t_{2})\\
 \intertext{and}
 \hat{O}_{s,2}(t,t^{\prime})=&\cos\omega(t-t^{\prime})\left\{\hat{x},\tilde{\rho}_{s}(t)\right\}
 -\frac{\sin\omega(t-t^{\prime})}{M\omega}\left\{\hat{p},\tilde{\rho}_{s}(t)\right\}\nonumber\\&
 -\frac{2}{M\omega}\int_{t^{\prime}}^{t}dt_{1}\int_{0}^{t}dt_{2}\sin\omega(t_{1}-t^{\prime})\alpha_{R}(t_{1}-t_{2})
 \hat{O}_{s,1}(t,t_{2})\nonumber\\&
 -\frac{2}{M\omega}\int_{t^{\prime}}^{t}dt_{1}\int_{0}^{t_{1}}dt_{2}\sin\omega(t_{1}-t^{\prime})\alpha_{I}(t_{1}-t_{2})
 \hat{O}_{s,2}(t,t_{2}).
\end{align}
In the derivation we have used the following functional derivatives
\begin{equation*}
M\left\{\frac{\delta \rho_{s}(t)}{\delta
\mu_{1}(t^{\prime})}+i\frac{\delta \rho_{s}(t)}{\delta
\mu_{4}(t^{\prime})}\right\}=\frac{2}{\sqrt{\hbar}}\int_{t^{\prime}}^tdt_{1}\alpha_{R}(t_{1}-t^{\prime})\hat{O}_{s,1}(t,t_{1})
\end{equation*}
and
\begin{equation*}M\left\{\frac{\delta\rho_{s}(t)}{\delta
\mu_{2}(t^{\prime})}-i\frac{\delta \rho_{s}(t)}{\delta
\mu_{3}(t^{\prime})}\right\}=\frac{2}{\sqrt{\hbar}}\int_{t^{'}}^tdt_{1}\alpha_{I}(t_{1}-t^{\prime})\hat{O}_{s,1}(t,t_{1}),
\end{equation*}
which are also readily obtained with the formal solution
of~$\rho_{s}(t)$~and the Furutsu-Novikov theorem.~For the integral
equations~(30)~and~(31) one can employ iteration to show that their
solutions~$\hat{O}_{s,1(2)}(t,t^{\prime})$~possess the following
forms~\begin{align}
\hat{O}_{s,1}(t,t^{\prime})=&x_{11}(t,t^{\prime})\left[\hat{p},\tilde{\rho}_{s}(t)\right]+x_{12}(t,t^{\prime})\left[\hat{x},\tilde{\rho}_{s}(t)\right]\\
\intertext{and}
\hat{O}_{s,2}(t,t^{\prime})=&x_{21}(t,t^{'})\left\{\hat{x},\tilde{\rho}_{s}(t)\right\}
+x_{22}(t,t^{\prime})\{\hat{p},\tilde{\rho}_{s}(t)\}+x_{23}(t,t^{\prime})\left[\hat{p},\tilde{\rho}_{s}(t)\right]
\nonumber\\&+x_{24}(t,t^{\prime})\left[\hat{x},\tilde{\rho}_{s}(t)\right].
\end{align}
Because the operators~$\left[\hat{x},\tilde{\rho}_{s}(t)\right]$,
$\left[\hat{p},\tilde{\rho}_{s}(t)\right]$,
$\left\{\hat{x},\tilde{\rho}_{s}(t)\right\}$,~and~$\{\hat{p},\tilde{\rho}_{s}(t)\}$
are arbitrary, their coefficients~$x_{jk}(t,t^{\prime})$~in
Eqs.~(32) and (33) are determined by Eqs.~(30) and~(31), which
satisfy the following integral equations
\begin{align}
x_{11}(t,t^{\prime})&=\frac{i}{M\omega}\sin\omega(t-t^{\prime})+\frac{2}{M\omega}\int_{t^{\prime}}^{t}
dt_{1}\int_{t_{1}}^{t}dt_{2}\sin\omega(t_{1}-t^{\prime})\alpha_{I}(t_{1}-t_{2})x_{11}(t,t_{2}),&\\
x_{12}(t,t^{\prime})&=-i\cos\omega(t-t^{\prime})+\frac{2}{M\omega}\int_{t^{\prime}}^{t}
dt_{1}\int_{t_{1}}^{t}dt_{2}\sin\omega(t_{1}-t^{\prime})\alpha_{I}(t_{1}-t_{2})x_{12}(t,t_{2}),&\\
x_{21}(t,t^{\prime})&=\cos\omega(t-t^{\prime})-\frac{2}{M\omega}\int_{t^{\prime}}^{t}dt_{1}
\int_{0}^{t_{1}}dt_{2}\sin\omega(t_{1}-t^{\prime})\alpha_{I}(t_{1}-t_{2})x_{21}(t,t_{2}),&\\
x_{22}(t,t^{\prime})&=-\frac{\sin\omega(t-t^{\prime})}{M\omega}-\frac{2}{M\omega}\int_{t^{\prime}}^{t}
dt_{1}\int_{0}^{t_{1}}dt_{2}\sin\omega(t_{1}-t^{\prime})\alpha_{I}(t_{1}-t_{2})x_{22}(t,t_{2}),&\\
x_{23}(t,t^{\prime})&=-\frac{2}{M\omega}\int_{t^{\prime}}^{t}dt_{1}\sin\omega(t_{1}-t^{\prime})
\left[\int_{0}^{t}dt_{2}\alpha_{R}(t_{1}-t_{2})x_{11}(t,t_{2})
+\int_{0}^{t_{1}}dt_{2}\alpha_{I}(t_{1}-t_{2})x_{23}(t,t_{2})\right],&\\
\nonumber\\ \intertext{and}
x_{24}(t,t^{\prime})&=-\frac{2}{M\omega}\int_{t^{\prime}}^{t}dt_{1}\sin\omega(t_{1}-t^{\prime})
\left[\int_{0}^{t}dt_{2}\alpha_{R}(t_{1}-t_{2})x_{12}(t,t_{2})
+\int_{0}^{t_{1}}dt_{2}\alpha_{I}(t_{1}-t_{2})x_{24}(t,t_{2})\right].
\end{align}
With these expressions at hand,~Eq.~(24) immediately becomes the
desired master equation, namely,
\begin{align}
i\hbar\frac{d\tilde{\rho}_{s}(t)}{dt}=&\left[\hat{H}_{s},\tilde{\rho}_{s}(t)\right]
+A_{1}(t)\left[\hat{x},\{\hat{x},\tilde{\rho}_{s}(t)\}\right]+A_{2}(t)\left[\hat{x},\left\{\hat{p},\tilde{\rho}_{s}(t)\right\}\right]
\nonumber\\&+A_{3}(t)\left[\hat{x},\left[\hat{p},\tilde{\rho}_{s}(t)\right]\right]+A_{4}(t)\left[\hat{x},\left[\hat{x},\tilde{\rho}_{s}(t)\right]\right],
\end{align}
where~$\hat{H}_{s}=\hat{p}^{2}/(2M)+M\omega^{2}\hat{x}^{2}/2$~and
the coefficients are
\begin{align}
A_{1}(t)&=\int_{0}^{t}dt^{\prime}
\alpha_{I}(t-t^{\prime})x_{21}(t,t^{\prime}),\\
A_{2}(t)&=\int_{0}^{t}dt^{\prime}
\alpha_{I}(t-t^{\prime})x_{22}(t,t^{\prime}),\\
A_{3}(t)&=\int_{0}^{t}dt^{\prime}\left[
\alpha_{R}(t-t^{\prime})x_{11}(t,t^{\prime})+
\alpha_{I}(t-t^{\prime})x_{23}(t,t^{\prime})\right],\\
\intertext{and} A_{4}(t)&=\int_{0}^{t}dt^{\prime}\left[
\alpha_{R}(t-t^{\prime})x_{12}(t,t^{\prime})+
\alpha_{I}(t-t^{\prime})x_{24}(t,t^{\prime})\right].
\end{align}
As clearly clarified in the literature~\cite{Hu,Yu},~$A_{1}(t)$~is
the coefficient for the frequency shift
because~$\left[\hat{x},\left\{\hat{x},\tilde{\rho}_{s}(t)\right\}\right]=\left[\hat{x}^{2},\tilde{\rho}_{s}(t)\right]$,
$A_{2}(t)$ is a quantum dissipation term, $A_{3}(t)$~reflects the
anomalous quantum diffusion while~$A_{4}(t)$~is the coefficient for
the normal quantum diffusion.
\section{equivalent to the Hu-Paz-Zhang equation}
\par The master equation and the dynamics of the dissipative harmonic oscillator has been derived and studied by several researchers with diversified
methods~\cite{Dekker,F.Haake,P.Schramm,Hu,Yu,Strunz,Chou,Chou1,
H.Grabert,Connell,Calzetta,Xu,peter,peter1,peter2,Fleming}.~Dekker
used the canonical quantization method~\cite{Dekker}, while Haake
and Reibold employed the Wigner function method. The latter also
studied the low-temperature and strong-damping
anomalies~\cite{F.Haake}. Grabert, Schramm, and Ingold went beyond
the factorized initial condition~\cite{P.Schramm}. It would be
stressed that the master equation in a general environment was
derived by Hu, Paz, and Zhang~(HPZ)~by virtue of path integral
technique~\cite{Hu}. An elementary derivation of HPZ equation was
documented by Halliwell and Yu with the Wigner function
approach~\cite{Yu}. Karrlein and Grabert pointed out that there is
an exact dissipative Liouville operator for certain correlated
initial conditions~\cite{H.Grabert}. The master equation was also
considered by Calzetta {\em{et al.}} through a stochastic method
based on the quantum Langevin equation~\cite{Calzetta}. Ford and
O'Connell~\cite{Connell} rederived the HPZ equation with the quantum
Langevin equation method and also analyzed its solution. Strunz and
Yu offered an alternative derivation, using the quantum trajectory
method~\cite{Strunz}. The HPZ master equation has recently been used
by Chou, Yu, and Hu to derive the master equation of two and more
coupled harmonic oscillators in a bosonic bath~\cite{Chou,Chou1}.
Furthermore, the HPZ master equation was extended by Xu {\em{et
al.}}~\cite{Xu} to the case where time dependent external fields are
applied.
\par For convenience of comparison, we will always use the results in the paper by Halliwell and Yu~\cite{Yu}. The HPZ master equation takes the
same form as Eq.~(40). The corresponding coefficients in our
notation read
\begin{align}
B_{1}(t)&=\int_{0}^{t}dt^{\prime}\alpha_{I}(t-t^{\prime})\bar{x}_{21}(t,t^{\prime}),\\
B_{2}(t)&=\int_{0}^{t}dt^{\prime}\alpha_{I}(t-t^{\prime})\bar{x}_{22}(t,t^{\prime}),\\
B_{3}(t)&=\int_{0}^{t}dt^{\prime}\left[\alpha_{R}(t-t^{\prime})\bar{x}_{11}(t,t^{\prime})+\alpha_{I}(t-t^{\prime})\bar{x}_{23}(t,t^{\prime})\right],\\
\intertext{and}
B_{4}(t)&=\int_{0}^{t}dt^{\prime}\left[\alpha_{R}(t-t^{\prime})\bar{x}_{12}(t,t^{\prime})+\alpha_{I}(t-t^{\prime})\bar{x}_{24}(t,t^{\prime})\right],
\end{align}
where
\begin{align}
\bar{x}_{21}(t,t^{\prime})&=u_{2}(t^{\prime})-\frac{\dot{u}_{2}(t)}{\dot{u}_{1}(t)}u_{1}(t^{\prime}),\\
\bar{x}_{22}(t,t^{\prime})&=\frac{u_{1}(t^{\prime})}{M\dot{u}_{1}(t)},\\
\bar{x}_{11}(t,t^{\prime})&=\frac{i}{M}G_{1}(t,t^{\prime}),\\
\bar{x}_{12}(t,t^{\prime})&=-iG^{\prime}_{1}(t,t^{\prime}),\\
\bar{x}_{23}(t,t^{\prime})&=\frac{2i}{M^{2}}\int_{t^{\prime}}^{t}dt_{1}\int_{0}^{t}dt_{2}\alpha_{R}(t_{1}-t_{2})G_{1}(t,t_{2})G_{2}(t^{\prime},t_{1}),\\
\nonumber \intertext{and}
\bar{x}_{24}(t,t^{\prime})&=-\frac{2i}{M}\int_{t^{\prime}}^{t}dt_{1}\int_{0}^{t}dt_{2}\alpha_{R}(t_{1}-t_{2})G^{\prime}_{1}(t,t_{2})G_{2}(t^{\prime},t_{1}).
\end{align} Here the dot over~$u_{j}(t)~(j=1,2)$~stands for
the derivative with respect to~$t$~and the functions~$u_{j}(t)$~are
the solutions of the homogeneous integro-differential equation
\begin{equation}
\left(\frac{d^{2}}{dt^{2}}+\omega^{2}\right)u(t)+\frac{2}{M}\int_{0}^{t}dt^{\prime}
\alpha_{I}(t-t^{\prime})u(t^{\prime})=0
\end{equation}
with inhomogeneous boundary conditions~$u_{1}(0)=1$,
$u_{1}(t)=0$~and~$u_{2}(0)=0$, $u_{2}(t)=1$~and~$G_{j}(t_{1},
t_{2})~(j=1,2)$~are the Green's functions obeying
\begin{equation}
\left(\frac{d^{2}}{dt_{1}^{2}}+\omega^{2}\right)G(t_{1},t_{2})+\frac{2}{M}\int_{0}^{t_{1}}dt_{3}
\alpha_{I}(t_{1}-t_{3}) G(t_{3},t_{2})=\delta(t_{1}-t_{2})
\end{equation}
with specified initial conditions at the fixed initial and final
times~$G_{1}(t_{1}=0, t_{2})=0$, $G_{1}^{\prime}(t_{1}=0, t_{2})=0$
and~$G_{2}(t_{1}=t, t_{2})=0$, $G_{2}^{\prime}(t_{1}=t, t_{2})=0$.
Here the prime in~$G^{\prime}_{j}(t_{1}, t_{2})$~stands for the
derivative with respect to the first variable, that is,
$G_{j}^{\prime}(t_{1}, t_{2})=\partial G_{j}(t_{1}, t_{2})/\partial
t_{1}$. Because of
causality~$G_{1}(t_{1},t_{2})=0$~for~$t_{1}<t_{2}$,
while~$G_{2}(t_{1},t_{2})=0$~for~$t_{1}>t_{2}$.
\par Now we show that
the HPZ equation and that derived with the stochastic formulation
are identical. To this end we only need to prove
that~$A_{j}(t)=B_{j}(t)$ ($j=1-4$), respectively. As clearly shown
in Eqs.~(41)$-$(44) and Eqs.~(45)$-$(48), all functions~$A_{j}(t)$
and the counterparts~$B_{j}(t)$~are integrals over the time
range~$[0, t]$. Therefore, a sufficient condition
for~$A_{j}(t)=B_{j}(t)$~is that the corresponding integrands are the
same. Besides, because these integrands consist of
factors~$\alpha_{R(I)}(t)$~that are dependent on the specificity of
the dissipation and can be arbitrary, one can further simplify the
proof significantly.
\subsection{Proof of $A_{1}(t)=B_{1}(t)$, $A_{2}(t)=B_{2}(t)$}
To prove~$A_{1}(t)=B_{1}(t)$, one should prove
$x_{21}(t,t^{\prime})=\bar{x}_{21}(t,t^{\prime})$. Note
that~$u_{j}(t)$~satisfy linear differential Eq.~(55).
Because~$\bar{x}_{21}(t,t^{\prime})$~is a linear combination
of~$u_{1}(t^{\prime})$~and~$u_{2}(t^{\prime})$, as a function
of~$t^{\prime}$, it should also obey Eq.~(55),
\begin{equation}
\left(\frac{\partial^{2}}{\partial
t^{\prime2}}+\omega^{2}\right)\bar{x}_{21}(t,t^{\prime})=
-\frac{2}{M}
\int_{0}^{t^{'}}dt_{1}\alpha_{I}(t^{\prime}-t_{1})\bar{x}_{21}(t,t_{1}).
\end{equation}
On the other hand, from the integral Eq.~(36) we can show by a
straightforward algebra that calculating the second order derivative
of~$x_{21}(t,t^{\prime})$~on both sides one can obtain the equation
the same as
Eq.~(57).~Moreover,~$x_{21}(t,t^{\prime})\mid_{t^{\prime}=t}=\bar{x}_{21}(t,t^{\prime})\mid_{t^{\prime}=t}=1$~and~$\partial
x_{21}(t,t^{\prime})/\partial t^{\prime}\mid_{t^{\prime}=t}=\partial
\bar{x}_{21}(t,t^{\prime})/\partial
t^{\prime}\mid_{t^{\prime}=t}=0$, the initial conditions are the
same. Therefore,~$A_{1}(t)$~is identical with~$B_{1}(t)$. On the
same line, one can prove~$A_{2}(t)=B_{2}(t)$.

\subsection{Proof of $A_{3}(t)=B_{3}(t)$, $A_{4}(t)=B_{4}(t)$}
As pointed out above, the problem of proving~$A_{3}(t)=B_{3}(t)$~can
be changed to proving the equivalence of the two involved
integrands. That is, one needs to show~$
x_{11}(t,t^{\prime})=\bar{x}_{11}(t,t^{\prime})$~and~
$x_{23}(t,t^{\prime})=\bar{x}_{23}(t,t^{\prime})$.
\par Let us look at Eq.~(56) with the initial condition~$G_{1}(0,t_{2})=0$~and~$G^{\prime}_{1}(0,t_{2})=0$.
Suppose the second term on the left hand side is given. Then
Eq.~(56) can be viewed as a function of~$t_{1}$~and can be ``solved"
with Green's function method. Now the required Green's function
obeys
\begin{equation*}
\left(\frac{d^{2}}{dt_{1}^{2}}+\omega^{2}\right)\bar{G}(t_{1},
\tau)=\delta(t_{1}-\tau)
\end{equation*}
with~$\bar{G}(t_{1},\tau)\mid_{t_{1}<\tau}=0$~and~$\partial\bar{G}(t_{1},\tau)/\partial
t_{1}\mid_{t_{1}<\tau}=0$. Its solution is
\begin{equation*}
\bar{G}(t_{1},\tau)=\frac{\sin\omega(t_{1}-\tau)}{\omega}\theta(t_{1}-\tau).
\end{equation*}
Therefore, the ``solution'' of Eq.~(56) can be written as
\begin{align*}
G_{1}(t_{1},t_{2})=&\int_{0}^{t_{1}}d\tau
\bar{G}(t_{1},\tau)\left[\delta(\tau-t_{2})-\frac{2}{M}\int_{t_{2}}^{\tau}dt_{4}\alpha_{I}(\tau-t_{4})G_{1}(t_{4},t_{2})\right]
\\=&\frac{\sin\omega(t_{1}-t_{2})}{\omega}-\frac{2}{M\omega}\int_{t_{2}}^{t_{1}}dt_{3}\int_{t_{2}}^{t_{3}}dt_{4}\sin\omega(t_{1}-t_{3})
\alpha_{I}(t_{3}-t_{4})G_{1}(t_{4},t_{2}).
\end{align*}
Substituting into Eq.~(51) yields
\begin{equation}
\bar{x}_{11}(t,t^{\prime})=\frac{i}{M\omega}\sin\omega(t-t^{\prime})-\frac{2}{M\omega}\int_{t^{\prime}}^{t}dt_{1}\int_{t^{\prime}}^{t_{1}}
dt_{2}\sin\omega(t-t_{1})\alpha_{I}(t_{1}-t_{2})\bar{x}_{11}(t_{2},t^{\prime}).
\end{equation}
Note that from Eq.~(20) one can
see~$\alpha_{I}(t_{1}-t_{2})=-\alpha_{I}(t_{2}-t_{1})$. Making
change of integration orders and variables in the double integral
then leads to
\begin{equation}
\bar{x}_{11}(t,t^{\prime})=\frac{i}{M\omega}\sin\omega(t-t^{\prime})+\frac{2}{M\omega}\int_{t^{\prime}}^{t}dt_{1}\int_{t_{1}}^{t}
dt_{2}\sin\omega(t-t_{2})\alpha_{I}(t_{1}-t_{2})\bar{x}_{11}(t_{1},t^{\prime}),
\end{equation}
which is an integral equation of~$\bar{x}_{11}(t,t^{\prime})$~with
respect to the first argument~$t$. Now we show
that~$\bar{x}_{11}(t,t^{\prime})$~determined by Eq.~(59) is
identical with~$x_{11}(t,t^{\prime})$~solved from Eq.~(34) that is
an integral equation with respect to the second
argument~$t^{\prime}$. To this end, we discretize the
variables~$t$~and~$t^{\prime}$~so
that~$x_{11}(t,t^{\prime})$~and~$\bar{x}_{11}(t,t^{\prime})$~can be
represented as matrices. To be specific, the elements of
matrices~$\emph{\textbf{X}}_{0}$, $\emph{\textbf{X}}$,
$\emph{\textbf{X}}^{\prime}$,  and~$\bm{\alpha}$~take the
discretized values of~$x_{0}(t,t^{\prime})$, $x_{11}(t,t^{\prime})$,
$\bar{x}_{11}(t,t^{\prime})$, and~$\alpha(t-t^{\prime})$,
respectively. Here~$x_{0}(t,t^{\prime})=\sin\omega(t-t^{\prime})$~is
introduced. With these matrices, the integrals in Eqs.~(34) and~(59)
become matrix products,
\begin{align}
\emph{\textbf{X}}&=\emph{\textbf{X}}_{0}-\emph{\textbf{X}}\emph{\bm{\alpha}}\emph{\textbf{X}}_{0}
\\\intertext{and}
\emph{\textbf{X}}^{\prime}&=\emph{\textbf{X}}_{0}-\emph{\textbf{X}}_{0}\emph{\bm{\alpha}}\emph{\textbf{X}}^{\prime}.
\end{align}
Solving these matrix equation with elementary algebraic
manipulations, we find \begin{align}
\emph{\textbf{X}}&=\emph{\textbf{X}}_{0}(\textbf{1}+\emph{\bm{\alpha}}\emph{\textbf{X}}_{0})^{-1}
\\\intertext{and}
\emph{\textbf{X}}^{\prime}&=\left(\textbf{1}+\emph{\textbf{X}}_{0}\emph{\bm{\alpha}}\right)^{-1}\emph{\textbf{X}}_{0}.\end{align}
Because
$\left(\textbf{1}+\emph{\textbf{X}}_{0}\emph{\bm{\alpha}}\right)\emph{\textbf{X}}_{0}=\emph{\textbf{X}}_{0}\left(\textbf{1}
+\emph{\bm{\alpha}}\emph{\textbf{X}}_{0}\right)$, one immediately
obtains~$\emph{\textbf{X}}=\emph{\textbf{X}}^{\prime}$. Therefore,
we find~$x_{11}(t,t^{\prime})=\bar{x}_{11}(t,t^{\prime})$. Thus,
Eq.~(51) can be recast as
~$G_{1}(t,t^{\prime})=-iMx_{11}(t,t^{\prime})$. Substituting into
Eq.~(53) leads to
\begin{align}
\bar{x}_{23}(t,t^{\prime})=\frac{2}{M}\int_{t^{\prime}}^{t}dt_{1}\int_{0}^{t}dt_{2}\alpha_{R}(t_{1}-t_{2})x_{11}(t,t_{2})G_{2}(t^{\prime},t_{1}).
\end{align}
We now treat~$G_{2}(t_{1},t_{2})$~in the same way as we did
for~$G_{1}(t_{1},t_{2})$~in the above. As a result,
$G_{2}(t_{1},t_{2})$~satisfies the following integral equation,
\begin{equation}
G_{2}(t_{1},t_{2})=\frac{\sin\omega(t_{1}-t_{2})}{\omega}-\frac{2}{M\omega}\int_{t_{1}}^{t}dt_{3}\int_{0}^{t_{3}}dt_{4}
\sin\omega(t_{3}-t_{1}) \alpha_{I}(t_{3}-t_{4})G_{2}(t_{4},t_{2}).
\end{equation}
Inserting into Eq.~(64) and rearranging, we find the integral
equation of~$\bar{x}_{23}(t,t^{\prime})$~is identical with that of
$x_{23}(t,t^{\prime})$,~Eq.~(38). Therefore,
$\bar{x}_{23}(t,t^{\prime})=x_{23}(t,t^{\prime})$. Similarly, we can
demonstrate the equality~$A_{4}(t)=B_{4}(t)$.
\section{master equation of driven harmonic oscillator}
Consider a dissipative harmonic oscillator driven by general
external time-dependent fields~\cite{Xu,peter,peter1,peter2} with
Hamiltonian~\begin{equation*} \hat{H}_{s}(t)
=\frac{\hat{p}^{2}}{2M}+\frac{1}{2}M\omega^{2}\hat{x}^{2}+f_{1}(t)\hat{x}+f_{2}(t)\hat{p}.
\end{equation*}
We shall work out the master equation in the driving case along the
same line of deriving the master equation of the dissipative
harmonic oscillator in Sec.~\Rmnum{3}. Note that for the external
time-dependent fields only act on the system, the bath-induced field
is the same as that without driving fields. We can thus start from
Eqs.~(26)~and~(27). By solving the equations of motion for the
Heisenberg operators and taking stochastic averaging, we find
\begin{align*}
\hat{O}_{s,1}(t,t^{\prime})=&-i\cos\omega\left(t-t^{\prime}\right)\left[\hat{x},\tilde{\rho}_{s}(t)\right]
+\frac{i}{M\omega}
\sin\omega\left(t-t^{\prime}\right)\left[\hat{p},\tilde{\rho}_{s}(t)\right]\\&
+\frac{2}{M\omega}\int_{t^{\prime}}^{t}dt_{1}\int_{t_{1}}^{t}dt_{2}\sin\omega(t_{1}-t^{\prime})\alpha_{I}
(t_{1}-t_{2})\hat{O}_{s,1}(t,t_{2})\\
\intertext{and}
\hat{O}_{s,2}(t,t^{\prime})=&\cos\omega(t-t^{\prime})\left\{\hat{x},\tilde{\rho}_{s}(t)\right\}-\frac{\sin\omega(t-t^{\prime})}{M\omega}\left\{\hat{p},\tilde{\rho}_{s}(t)\right\}\\&
-\frac{2}{M\omega}\int_{t^{\prime}}^{t}dt_{1}\int_{0}^{t}dt_{2}\sin\omega(t_{1}-t^{\prime})\alpha_{R}(t_{1}-t_{2})\hat{O}_{s,1}(t,t_{2})\\&
-\frac{2}{M\omega}\int_{t^{\prime}}^{t}dt_{1}\int_{0}^{t_{1}}dt_{2}\sin\omega(t_{1}-t^{\prime})\alpha_{I}(t_{1}-t_{2})\hat{O}_{s,2}(t,t_{2})
\\&-\frac{2}{M\omega}\int_{t^{\prime}}^{t}dt_{1}\sin\omega\left(t_{1}-t^{\prime}\right)f_{1}(t_{1})\tilde{\rho}_{s}(t)-2\int_{t^{\prime}}^{t}dt_{1}\cos\omega(t_{1}-t^{\prime})f_{2}(t_{1})\tilde{\rho}_{s}(t).
\end{align*}
Similar to the undriven case, these dissipation operators possess
the following forms
\begin{align*}
\hat{O}_{s,1}(t,t^{\prime})=&x_{11}(t,t^{\prime})\left[\hat{p},\tilde{\rho}_{s}(t)\right]
+x_{12}(t,t^{\prime})\left[\hat{x},\tilde{\rho}_{s}(t)\right]\\
\intertext{and}
\hat{O}_{s,2}(t,t^{\prime})=&x_{21}(t,t^{\prime})\left\{\hat{x},\tilde{\rho}_{s}(t)\right\}+
x_{22}(t,t^{\prime})\left\{\hat{p},\tilde{\rho}_{s}(t)\right\}+x_{23}(t,t^{\prime})\left[\hat{p},\tilde{\rho}_{s}(t)\right]+x_{24}(t,t^{\prime})\left[\hat{x},\tilde{\rho}_{s}(t)\right]
\\&+x_{25}(t,t^{\prime})\tilde{\rho}_{s}(t),
\end{align*}
where all
coefficients~$x_{jk}(t,t^{\prime})$~except~$x_{25}(t,t^{\prime})$~are
the same as that of undriven case Eqs.~(34)$-$(39). The new term is
determined by
\begin{align*}
x_{25}(t,t^{\prime})=&-\frac{2}{M\omega}\int_{t^{\prime}}^{t}dt_{1}\sin\omega(t_{1}-t^{\prime})f_{1}(t_{1})-2\int_{t^{\prime}}^{t}dt_{1}\cos\omega(t_{1}-t^{\prime})f_{2}(t_{1})
\\&-\frac{2}{M\omega}\int_{t^{\prime}}^{t}dt_{1}\int_{0}^{t_{1}}dt_{2}\sin\omega(t_{1}-t^{\prime})\alpha_{I}(t_{1}-t_{2})x_{25}(t,t_{2}).
\end{align*}
Therefore, substituting
$M\{\rho_{s,1(2)}(t,t^{\prime})\}$~by~$\hat{O}_{s,1(2)}(t,t^{\prime})$
in Eq.~(24) yields the required master equation
\begin{align*}
i\hbar\frac{d
\tilde{\rho_{s}}(t)}{dt}=&\left[\hat{H}_{eff}(t),\tilde{\rho_{s}}(t)\right]
+A_{1}(t)\left[\hat{x},\left\{\hat{x},\tilde{\rho}_{s}(t)\right\}\right]
+A_{2}(t)\left[\hat{x},\left\{\hat{p},\tilde{\rho_{s}}(t)\right\}\right]
\\&+A_{3}(t)\left[\hat{x},\left[\hat{p},\tilde{\rho_{s}}(t)\right]\right]+A_{4}(t)\left[\hat{x},\left[\hat{x},\tilde{\rho_{s}}(t)\right]\right],
\end{align*}
where the effective Hamiltonian~$\hat{H}_{eff}(t)$~is
\begin{equation*}
\hat{H}_{eff}(t)=\hat{H}_{s}(t)+\hat{x}\int_{0}^{t}dt^{\prime}\alpha_{I}(t-t^{\prime})x_{25}(t,t^{\prime}).
\end{equation*} Here again, the coefficients~$A_{j}(t)$~$(j=1-4)$~are the same as
the undriven case, which are given through Eqs.~(41)$-$(44). It is
obvious that the second term in~$\hat{H}_{eff}(t)$~reflects the very
interplay between the system and the bath mediated by the driving
fields.
\section{Conclusion}
Classical dynamics of dissipative systems is traditionally described
by the Langevin equation. It has been shown that the stochastic
formulation provides a similar description to quantum dissipative
systems~\cite{Shao}. In this framework the dissipative system obeys
Liouville equations subjected to the stochastic fields due to the
bath. Based on the stochastic formulation, flexible numerical
methods have been proposed and used to solve dissipative
dynamics~\cite{Denis}. As complementing to previous work the present
paper provides conceptual and analytical results. We have
illustrated how to obtain the bath-induced field of the
Caldeira-Leggett model through elementary solution of quantum linear
systems. Furthermore, we have elaborated a systematic approach to
derive the master equation, if it exists.
\par The reduced density matrix can in principle be obtained by
solving the stochastic Liouville equation and calculating stochastic
average. Because it is difficult to have convergent stochastic
averaging, a master equation describing the evolution of the reduced
density matrix is highly desired. We have shown the existence of the
master equation relies not only on the feature of the dissipation
charactered by the spectral density function, but also on the
dynamics of the stochastic system itself. For linear systems, we
have found that the ``dissipative operator" due to the interplay of
the system and the stochastic field is exactly solvable and thereby
derived the master equation. We have shown that the master equation
is equivalent to the HPZ equation derived by Hu, Paz, and Zhang
using path-integral approach~\cite{Hu,Yu}. We would like to point
out that the coefficients in our master equation are determined by a
set of integral equations, which may not suffer from the
mathematical problems concerned by Fleming, Roura, and
Hu~\cite{Fleming}.\par We have also shown that the master equation
of driven harmonic oscillator can be derived similarly. In this
case, the system is dressed by both the driving and the stochastic
fields, although the dissipation operators appear the same as that
of the undriven case.
\section*{Acknowledgments}
This work is supported by the National Natural Science Foundation of
China (No.~91027013) and the 973 program of the Ministry of Science
and Technology of China (2011CB808502 and 2007CB815206).

\end{document}